\documentstyle[12pt,prd,aps,epsfig,preprint]{revtex}
\begin{document}
\draft
\date{\today}
\title{Scalar $a_0(980)$ meson in $\phi\rightarrow \pi^{0}\eta\gamma$ decay}

\author{A. Gokalp~\thanks{agokalp@metu.edu.tr}, A. K\"{u}\c{c}\"{u}karslan, S. Solmaz and O. Yilmaz~\thanks{oyilmaz@metu.edu.tr}}
\address{ {\it Physics Department, Middle East Technical University,
06531 Ankara, Turkey}}
\maketitle

\begin{abstract}
We study the radiative decay $\phi\rightarrow\pi^{0}\eta\gamma$
within the framework of a phenomenological approach in which the
contributions of $\rho$-meson, chiral loop and $a_0$-meson are
considered. We analyze the interference effects between different
contributions and utilizing the experimental branching ratio and
invariant $\pi^0\eta$ mass spectrum for
$\phi\rightarrow\pi^{0}\eta\gamma$ decay we estimate the branching
ratio of $\phi\rightarrow a_0\gamma$ decay.
\end{abstract}

\thispagestyle{empty} ~~~~\\ \pacs{PACS numbers: 12.20.Ds,
13.40.Hq, 14.40.-n }
%\narrowtext
\newpage
\setcounter{page}{1}
%%%
%%%
\section{Introduction}
The nature and quark substructure of low mass scalar mesons, in
particular those of isoscalar $f_0(980)$ and isovector $a_0(980)$,
have been a subject of controversy in hadron spectroscopy over the
years. A variety of interpretations have been proposed for their
structure, but, whether they are conventional $q\bar{q}$ states in
quark model \cite{R1}, $K\bar{K}$ molecules \cite{R2}, or exotic
multiquark $q^2\bar{q}^2$ states \cite{R3} have not been
established yet.

It was suggested by Achasov and Ivanchenko \cite{R4} that the
radiative decays of $\phi$ meson to pseudoscalar mesons,
$\phi\rightarrow \pi^0\pi^0\gamma$ and $\phi\rightarrow
\pi^0\eta\gamma$, offer the possibility of obtaining information
on the nature of $f_0(980)$ and $a_0(980)$ mesons, respectively.
Close, Isgur and Kumano \cite{R5} noted that the radiative decays
$\phi\rightarrow S\gamma$, where $S=f_0$ or $a_0$, can be utilized
to differentiate among various models of the structure of these
scalar mesons. They shown that although the transition rates
$\Gamma(\phi\rightarrow f_0\gamma)$ and $\Gamma(\phi\rightarrow
a_0\gamma)$ depends on the unknown dynamics, the ratio of the
decay rates $\Gamma(\phi\rightarrow
f_0\gamma)/\Gamma(\phi\rightarrow a_0\gamma)$ will be sensitive to
the spatial distribution of quarks and the spatial wavefunctions
of scalar mesons, and thus it provides an experimental test which
distinguishes between alternative explanations of their structure.
In both the $K\bar{K}$ molecule, and four-quark cluster
$q^2\bar{q}^2$ pictures of the structure of scalar mesons, it is
now generally accepted that the radiative decays $\phi\rightarrow
S\gamma$ proceed by the mechanism in which $\phi$ and $S$ both
couple to an intermediate $K^+K^-$ loop in the chain of reactions
$\phi\rightarrow K^+K^- \rightarrow K^+K^-\gamma\rightarrow
S\gamma$ \cite{R4,R5}. The corresponding decay rates were obtained
as \cite{R5}
\begin{equation}\label{e1}
BR(\phi\rightarrow f_0(980)\gamma)=BR(\phi\rightarrow
a_0(980)\gamma)\simeq (2.0\pm 0.5)\times 10^{-4}\times F^2(R)~~,
\end{equation}
where $F^2(R)$ is a form factor that depends on the spatial
wavefunctions of the scalar mesons, and $F^2(R)=1$ in point-like
effective field theory calculations. In obtaining this result it
is further assumed that both $f_0(980)$ and $a_0(980)$ are isospin
eigenstates, and $g^2_{SK^+K^-}/4\pi=0.58~GeV^2$. If $f_0(980)$
and $a_0(980)$ are spatially extended $K\bar{K}$ molecules with
$R>\Lambda^{-1}_{QCD}$, then $F^2(R)<1$ and the resulting
branching ratios are BR($\phi\rightarrow
S\gamma)\simeq(0.4-1)\times 10^{-4}$ \cite{R5}. If the meson
states are compact four-quark clusters confined within
$R\sim\Lambda^{-1}_{QCD}$ like in $q^2\bar{q}^2$ picture, then the
resulting branching ratios are expected as BR($\phi\rightarrow
S\gamma)\simeq 2\times 10^{-4}$ \cite{R4,R5}. In addition to the
predictions about the absolute branching ratios of
$\phi\rightarrow S\gamma$ decays, there is also the weaker result
that follows from Eq. 1, that is theoretically $BR(\phi\rightarrow
f_0 \gamma)/BR(\phi\rightarrow a_0 \gamma)=1$.

The very accurate data that have been obtained from Novosibirsk
SND \cite{R6,R7} and CMD-2 \cite{R8} Collaborations give the
following branching ratios for $\phi\rightarrow \pi^0\pi^0\gamma$
and $\phi\rightarrow\pi^{0}\eta\gamma$  decays:
BR($\phi\rightarrow\pi^{0}\pi^0\gamma)=(1.221\pm 0.098\pm
0.061)\times 10^{-4}$ \cite{R6},
BR($\phi\rightarrow\pi^{0}\eta\gamma)=(0.88\pm 0.14\pm 0.09)\times
10^{-4}$ \cite{R7}, and
BR($\phi\rightarrow\pi^{0}\pi^0\gamma)=(0.92\pm 0.08\pm
0.06)\times 10^{-4}$,
BR($\phi\rightarrow\pi^{0}\eta\gamma)=(0.90\pm 0.24\pm 0.10)\times
10^{-4}$ \cite{R8}, where the first error is statistical and the
second one is systematic. From the analysis of their experimental
results, Achasov et al. \cite{R6} concluded that the $f_0\gamma$
mechanism dominates the $\phi\rightarrow\pi^{0}\pi^0\gamma$ decay
and the contributions coming from $\sigma\gamma$ and $\rho^0\pi^0$
intermediate states are small. Neglecting these contributions they
obtained the branching ratio BR($\phi\rightarrow
f_0\gamma)=(3.5\pm 0.3^{+1.3}_{-0.5})\times 10^{-4}$, where the
second errors are systematic and includes uncertainty in the
interference terms. This branching ratio was also obtained by
Akhmetshin et al. from their data of
$\phi\rightarrow\pi^{0}\pi^0\gamma$ decay as BR($\phi\rightarrow
f_0(980)\gamma)=(2.90\pm 0.21\pm 1.54)\times 10^{-4}$ \cite{R8}.
The $\phi\rightarrow f_0(980)\gamma$ decay branching ratio is now
quoted as an average of these two results which is
BR($\phi\rightarrow f_0(980)\gamma)=(3.4\pm 0.4)\times 10^{-4}$
\cite{R9}. Achasov et al., on the other hand, by assuming that
$a_0\gamma$ intermediate state dominates the
$\phi\rightarrow\pi^{0}\eta\gamma$ decay, and the contributions
from other decay mechanisms, for example
$\phi\rightarrow\rho^0\pi^0$, $\rho^0\rightarrow\eta\gamma$, can
be neglected, obtained the value for the branching ratio of the
decay $\phi\rightarrow a_0(980)\gamma$ as BR($\phi\rightarrow
a_0(980)\gamma)=(0.88\pm 0.17)\times 10^{-4}$ from the data of
their $\phi\rightarrow\pi^{0}\eta\gamma$ decay  experiment
\cite{R8}. Therefore, these experimental results suggest for the
ratio of the decay rates of the decays $\phi\rightarrow
f_0(980)\gamma$ and $\phi\rightarrow a_0(980)\gamma$ the
experimental value $\Gamma(\phi\rightarrow f_0
\gamma)/\Gamma(\phi\rightarrow a_0 \gamma)=3.8\pm 1.2$, with which
the theoretical result $\Gamma(\phi\rightarrow f_0
\gamma)/\Gamma(\phi\rightarrow a_0 \gamma)\simeq 1$ is not in
agreement.

In this paper, we study the role of the scalar $a_0(980)$ meson in
the mechanism of the radiative  $\phi\rightarrow\pi^{0}\eta\gamma$
decay employing a phenomenological framework, in which we try to
assess the roles of different processes and the contributions to
the decay rate coming from their amplitude in the mechanism of
this decay, and this way we attempt to estimate the branching
ratio BR($\phi\rightarrow a_0(980)\gamma)$ of the decay
$\phi\rightarrow a_0(980)\gamma$ from the experimental data of the
radiative $\phi\rightarrow\pi^{0}\eta\gamma$ decay. In our
analysis, we use the experimental data of Novosibirsk SND
Collaboration \cite{R7}.

\section{Formalism}

The radiative decay process of the type $V^0\rightarrow
P^{0}P^{0}\gamma$ where V and P belong to the lowest multiplets of
vector (V) and pseudoscalar (P) mesons have been studied using
different approaches. In particular, the branching ratio
$BR(\phi\rightarrow\pi^{0}\eta\gamma)$ has been calculated as
$BR(\phi\rightarrow\pi^{0}\eta\gamma)_{VDM}=5.4\times 10^{-6}$
\cite{R10} by considering intermediate vector meson contribution
only, and as $BR(\phi\rightarrow\pi^{0}\eta\gamma)_\chi=3\times
10^{-5}$ \cite{R11} by using chiral loop model within the
framework of chiral phenomenological Lagrangians.

In order to include the scalar $a_0$ meson resonance pole in the
decay mechanism for the calculation of the decay rate of
$\phi\rightarrow\pi^{0}\eta\gamma$ decay in a phenomenological
framework we consider two different approaches. In both
approaches, we do not make any assumptions about the structure of
the scalar $a_0$ meson.

In the first approach, which we name Model I, we include the
scalar $a_0$ resonance in an ad hoc manner. In this approach, we
assume that the mechanism of the
$\phi\rightarrow\pi^{0}\eta\gamma$ decay consists of the reactions
shown by the diagrams in Fig. 1. We describe the $\phi a_0\gamma$
and  $a_0\pi^0\eta$ vertices in Fig. 1(c) by the phenomenological
Lagrangians
\begin{eqnarray}\label{e2}
{\cal L}_{\phi a_0\gamma}=\frac{e}{M_\phi}g_{\phi
a_0\gamma}[\partial^\alpha \phi^\beta\partial_\alpha
A_\beta-\partial^\alpha \phi^\beta\partial_\beta A_\alpha]a_0
\end{eqnarray}
and
\begin{eqnarray}\label{e3}
{\cal L}_{a_0 \pi\eta}=g_{a_0\pi\eta}\vec{\pi}\cdot\vec{\pi}a_0~~,
\end{eqnarray}
respectively, which also serve to define the coupling constants
$g_{\phi a_0\gamma}$ and $g_{a_0\pi\eta}$. Since there are no
direct experimental results relating to $\phi a_0\gamma$-vertex,
but only an upper limit for the branching ratio of the decay
$\phi\rightarrow a_0\gamma$ is quoted in Review of Particle
Properties as $BR(\phi\rightarrow a_0 \gamma)<5\times 10^{-3}$
\cite{R9}, we determine this coupling constant in our calculation
by employing the experimental branching ratio of the
$\phi\rightarrow\pi^{0}\eta\gamma$ decay. The decay rates for the
$\phi\rightarrow a_0\gamma$ and the $a_0\rightarrow\pi^0\eta$
decays resulting from the above Lagrangians are
\begin{eqnarray}\label{e4}
\Gamma (\phi\rightarrow a_0\gamma)=
\frac{\alpha}{24\pi}\frac{(M_\phi^2-M_{a_0}^2)^3}{M_{\phi}^5}
g_{\phi a_0\gamma}^2~~,
\end{eqnarray}
and
\begin{eqnarray}\label{e5}
\Gamma (a_0\rightarrow \pi^0\eta)= \frac{g_{ a_0\pi\eta}^2}{16\pi
M_{a_0}
}\sqrt{\left[1-\frac{(M_{\pi^0}+M_\eta)^2}{M_{a_0}^2}\right]
\left[1-\frac{(M_{\pi^0}-M_\eta)^2}{M_{a_0}^2}\right]}~~,
\end{eqnarray}
respectively. We use the value $\Gamma_{a_0}=(0.069\pm 0.011)$ GeV
determined by E852 collaboration at BNL \cite{R12}, and we obtain
the coupling constant $g_{ a_0\pi\eta}$ as $g_{
a_0\pi\eta}=(2.32\pm 0.18)$ GeV.

The $\phi\rho\pi$ vertex in Fig. 1(a) is conventionally described
by the phenomenological Lagrangian
\begin{eqnarray}\label{e6}
{\cal
L}_{\phi\rho\pi}=g_{\phi\rho\pi}\epsilon^{\mu\nu\alpha\beta}\partial_\mu
\phi_\nu\partial_\alpha \rho_\beta\pi~~,
\end{eqnarray}
and the coupling constant $g_{\phi\rho\pi}$ was determined by
Achasov and Gubin using the data on the decay
$\phi\rightarrow\rho\pi\rightarrow\pi^+\pi^-\pi^0$ \cite{R9} as
$g_{\phi\rho\pi}=(0.811\pm 0.081)~~GeV^{-1}$ \cite{R13}. The
$\rho\pi\gamma$ vertex in Fig. 1(a) is described by the
phenomenological Lagrangian
\begin{eqnarray}\label{e7}
{\cal
L}_{\rho\eta\gamma}=g_{\rho\eta\gamma}\epsilon^{\mu\nu\alpha\beta}\partial_\mu
\rho_\nu\partial_\alpha A_\beta\eta~~,
\end{eqnarray}
and the coupling constant $g_{\rho\eta\gamma}$ is then obtained
from the experimental partial width of the radiative decay
$\rho\rightarrow\eta\gamma$ \cite{R9} as
$g_{\rho\eta\gamma}=(0.45\pm 0.07)~GeV^{-1}$.

We describe the $\phi KK$ vertex in Fig. 1(b) by the
phenomenological Lagrangian
\begin{eqnarray}\label{e8}
{\cal L}_{\phi K^+K^-}=-ig_{\phi KK} \phi^\mu (
   K^{+}\partial_{\mu}K^{-}-K^{-}\partial_{\mu}K^{+})
\end{eqnarray}
which results from the standard chiral Lagrangians in the lowest
order of chiral perturbation theory \cite{R14}. The decay rate for
the $\phi\rightarrow K^+K^-$ decay resulting from this Lagrangian
is
\begin{eqnarray}\label{e9}
\Gamma (\phi\rightarrow K^{+}K^{-})= \frac{g^{2}_{\phi
KK}}{48\pi}M_{\phi} \left [
1-\left(\frac{2M_{K}}{M_{\phi}}\right)^{2}\right ] ^{3/2}~~.
\end{eqnarray}
We utilize the experimental value for the branching ratio
$BR(\phi\rightarrow K^{+}K^{-})=(0.492\pm0.007)$ for the decay
$\phi\rightarrow K^{+}K^{-}$ \cite{R9}, and determine the coupling
constant $g_{\phi KK}$ as $g_{\phi KK}=(4.59\pm 0.05)$. For the
four pseudoscalar $KK\pi\eta$ amplitude, we use the result
obtained in standard chiral perturbation theory \cite{R15} which
is
\begin{eqnarray}\label{e10}
{\cal M} (K^{+}K^{-}\rightarrow\pi^0\eta)=
\frac{\sqrt{3}}{4f_\pi^2}
\left(M_{\pi^0\eta}^2-\frac{4}{3}M_{K}^2\right)~~.
\end{eqnarray}
where $\eta-\eta^\prime$ mixing is neglected, $M_{\pi^0\eta}$ is
the invariant mass of the $\pi^0\eta$ system, and we use
$f_\pi=92.4$ MeV. We therefore obtain the amplitude for the
diagram in Fig. 1(b) as
\begin{eqnarray}\label{e11}
{\cal M}=-\frac{e~g_{\phi KK}}{2\pi^{2}iM_{K}^{2}}\left[(p\cdot
k)(\epsilon\cdot u)-(p\cdot\epsilon)(k\cdot u)\right] I(a,b) {\cal
M} (K^{+}K^{-}\rightarrow\pi^0\eta)
\end{eqnarray}
where $(u,p)$ and $(\epsilon, k)$ are the polarizations and
four-momenta of the $\phi$ meson and the photon, respectively, and
$a=M_{\phi}^{2}/M_{K}^{2}$, $b=M_{\pi^0\eta}^{2}/M_{K}^{2}$ with
the invariant mass of the final $\pi^0\eta$ system given by
$M_{\pi^0\eta}^2=(q_1+q_2)^2=(p-k)^2$. The loop function is
defined as \cite{R16,R17}
\begin{eqnarray}\label{e12}
I(a,b)=\frac{1}{2(a-b)} -\frac{2}{(a-b)^{2}}\left [
f\left(\frac{1}{b}\right)-f\left(\frac{1}{a}\right)\right ]
+\frac{a}{(a-b)^{2}}\left [
g\left(\frac{1}{b}\right)-g\left(\frac{1}{a}\right)\right ]
\end{eqnarray}
where
\begin{eqnarray}\label{e13}
&&f(x)=\left \{ \begin{array}{rr}
           -\left [ \arcsin (\frac{1}{2\sqrt{x}})\right ]^{2}~,& ~~x>\frac{1}{4} \\
\frac{1}{4}\left [ \ln (\frac{\eta_{+}}{\eta_{-}})-i\pi\right
]^{2}~, & ~~x<\frac{1}{4}
            \end{array} \right.
\nonumber \\ && \nonumber \\ &&g(x)=\left \{ \begin{array}{rr}
        (4x-1)^{\frac{1}{2}} \arcsin(\frac{1}{2\sqrt{x}})~, & ~~ x>\frac{1}{4} \\
 \frac{1}{2}(1-4x)^{\frac{1}{2}}\left [\ln (\frac{\eta_{+}}{\eta_{-}})-i\pi \right ]~, & ~~ x<\frac{1}{4}
            \end{array} \right.
\nonumber \\ && \nonumber \\ &&\eta_{\pm}=\frac{1}{2x}\left [
1\pm(1-4x)^{\frac{1}{2}}\right ] ~.
\end{eqnarray}

On the other hand, the introduction of the $a_0$ amplitude as in
Fig. 1(c) may be considered not to be very realistic since this
diagram implies direct quark transition and it thus makes a very
small contribution because of OZI supression. Indeed, it has been
shown that the scalar resonances $f_0(980)$ and $a_0(980)$ can be
excited from the chiral loops, with the loop iteration provided by
the Bethe-Salpeter equation using a kernel from the lowest order
chiral Lagrangian \cite{R18}. Furthermore, it has been argued that
the experimental data obtained in Novosibirsk give resonable
arguments in favour of the one-loop mechanism for $\phi\rightarrow
K^+K^-\rightarrow a_0\gamma$ and $\phi\rightarrow
K^+K^-\rightarrow f_0\gamma$ decays \cite{R19}. Therefore, in our
second approach to study the radiative
$\phi\rightarrow\pi^{0}\eta\gamma$ decay, which we name Model II,
we assume that this decay proceeds through the reactions
$\phi\rightarrow \rho^0\pi^0\rightarrow\pi^{0}\eta\gamma$,
$\phi\rightarrow K^+K^-\gamma\rightarrow\pi^{0}\eta\gamma$, and
$\phi\rightarrow a_0\gamma\rightarrow\pi^{0}\eta\gamma$ where the
last reaction proceeds by a two-step mechanisms with $a_0$
coupling to $\phi$ with intermediate $K\bar{K}$ states. We show
the processes contributing to the $\phi\rightarrow
\pi^{0}\eta\gamma$ decay amplitude diagramatically in Fig. 2. We
note that we do not make any assumptions about the structure of
the $a_0$ meson, and only assume that the $\phi$ and $a_0$ mesons
both couple strongly to the $K^+K^-$ system, as a result of which
there is an amplitude for the decay $\phi\rightarrow a_0\gamma$ to
proceed through the charged kaon loop independent of the nature
and the dynamical structure of $a_0$ meson. The only new
ingredient required in our second approach is the $K^+K^-a_0$
vertex which we assume is described by the phenomenological
Lagrangian
\begin{eqnarray}\label{e14}
{\cal L}_{a_0 K^+K^-}=g_{a_0 K^+K^-}~ K^{+}K^{-}a_0~~.
\end{eqnarray}
The decay width of $a_0$ that follows from this Lagrangian is
\begin{eqnarray}\label{e15}
\Gamma(a_0\rightarrow K^+K^-)=\frac{g_{a_0K^+K^-}^2}{16\pi
M_{a_0}}\left[1-\left(\frac{2M_K}{M_{a_0}}\right)^2\right]^{1/2}
\end{eqnarray}
which is usually considered to define the coupling constant
$g_{a_0K^+K^-}$. In our second approach, we calculate the decay
rate of $\phi\rightarrow \pi^{0}\eta\gamma$  decay using the
diagrams shown Fig. 2, and by utilizing the experimental value of
this decay rate determine the coupling constant $g_{a_0K^+K^-}$.

In our calculation of the invariant amplitudes we make the
replacement $p^2-M^2\rightarrow p^2-M^2+iM\Gamma$ in $a_0$ and
$\rho^0$ propagators and use the experimental value
$\Gamma_{\rho}=(150.2\pm0.8)$ MeV \cite{R9} for $\rho^0$ meson,
because using a $q^2$-dependent width did not affect our results
appreciably. However, since the mass of the $M_{a_0}$ meson is
very close to the $K^+K^-$ threshold, this induces a strong energy
dependence in the width of the $a_0$-meson. We follow the widely
accepted option to deal with this problem that was proposed by
Flatt$\acute{e}$ \cite{R20} based on a coupled channel
$(\pi\eta,K\bar{K})$ description of the $a_0$ resonance, and
parametrize the $a_0$ width as
\begin{eqnarray}\label{e16}
  \Gamma^{a_0}(q^2)&=&\Gamma^{a_0}_{\pi^0\eta}(q^2)
  ~\theta(\sqrt{q^2}-(M_{\pi^0}+M_\eta))
  \nonumber \\&&+
ig_{K\overline{K}}\sqrt{M_K^2-q^2/4}~\theta(2M_K-\sqrt{q^2})
+g_{K\overline{K}}\sqrt{q^2/4-M_K^2}~\theta(\sqrt{q^2}-2M_K)~~,
 \end{eqnarray}
where
\begin{equation}\label{e17}
  \Gamma^{a_0}_{\pi^0\eta}(q^2)=\frac{g_{a_0\pi\eta}^2}{16\pi(q^2)^{3/2}}
  ~\sqrt{[q^2-(M_{\pi^0}+M_\eta)^2][q^2-(M_{\pi^0}-M_\eta)^2]}~~.
\end{equation}
We use the Flatt$\acute{e}$ parameter $g_{K\bar{K}}$ as
$g_{K\bar{K}}=(0.22\pm 0.04)$ which was determined by E852
Collaboration at BNL by a fit to data in their experiment in which
they determined the parameters of the $a_0$ meson \cite{R12}. In
all our calculations and in the analysis of the experimental
invariant $\pi^0\eta$ mass spectrum of the $\phi\rightarrow
\pi^{0}\eta\gamma$ decay, therefore, for the mass of $a_0$ meson
we use the value $M_{a_0}=(0.991\pm 0.0025)$ GeV as determined in
this experiment by E852 Collaboration \cite{R12}.

We express  the invariant amplitude ${\cal M}$(E$_{\gamma}, E_1)$
for the decay $\phi\rightarrow\pi^{0}\eta\gamma$ as ${\cal M}=
{\cal M}_a+ {\cal M}_b+ {\cal M}_c$  in Model I and as ${\cal M}=
{\cal M}_a'+ {\cal M}_b'+ {\cal M}_c'$ in Model II, where ${\cal
M}_a$, ${\cal M}_b$, and ${\cal M}_c$ are the invariant amplitudes
resulting from the diagrams (a), (b), and (c) in Fig. 1
respectively, describing the Model I, and ${\cal M}_a'$, ${\cal
M}_b'$ and ${\cal M}_c'$ are the invariant amplitudes
corresponding to the diagrams (a), (b) and (c) in Fig. 2,
respectively, defining the Model II. This way we take the
interference between different reactions contributing to the decay
$\phi\rightarrow\pi^{0}\eta\gamma$ into account. Then the
differential decay probability for
$\phi\rightarrow\pi^{0}\eta\gamma$ for an unpolarized $\phi$ meson
at rest is given as
\begin{eqnarray}\label{e18}
\frac{d\Gamma}{dE_{\gamma}dE_{1}}=\frac{1}{(2\pi)^{3}}~\frac{1}{8M_{\phi}}~
\mid {\cal M}\mid^{2} ,
\end{eqnarray}
where E$_{\gamma}$ and E$_{1}$ are the photon and pion energies
respectively. We perform an average over the spin states of $\phi$
meson and a sum over the polarization states of the photon. The
decay width $\Gamma(\phi\rightarrow\pi^{0}\eta\gamma)$ is then
obtained by integration
\begin{eqnarray}\label{e19}
\Gamma=\int_{E_{\gamma,min.}}^{E_{\gamma,max.}}dE_{\gamma}
       \int_{E_{1,min.}}^{E_{1,max.}}dE_{1}\frac{d\Gamma}{dE_{\gamma}dE_{1}}
\end{eqnarray}
where the minimum photon energy is E$_{\gamma, min.}=0$ and the
maximum photon energy is given as
$E_{\gamma,max.}=[M_{\phi}^{2}-(M_{\pi^0}+M_\eta)^{2}]/2M_{\phi}$.
The maximum and minimum values for pion energy E$_{1}$ are given
by
\begin{eqnarray}\label{e19a}
\frac{1}{2(2E_{\gamma}M_\phi-M_\phi^{2})} [
-2E_{\gamma}^{2}M_\phi-M_\phi(M_\phi^2+M_{\pi^0}^2-M_\eta^2)+E_{\gamma}(3M_\phi^{2}+M_{\pi^0}^2-M_\eta^2)~~~~~~~~~~~~~~
\nonumber \\ \pm E_{\gamma}
\sqrt{4E_{\gamma}^2M_\phi^2+M_\phi^4+(M_{\pi^0}^2-M_\eta^2)^2-2M_\phi^2(M_{\pi^0}^2+M_\eta^2)+4E_{\gamma}
M_\phi(-M_\phi^{2}+M_{\pi^0}^2+M_\eta^2)}~] . \nonumber
\end{eqnarray}

\section{Results and Discussion}
In order to determine the coupling constants $g_{\phi a_0\gamma}$
in Model I and $g_{a_0 K^+K^-}$ in Model II, we use the
experimental value of the branching ratio for the radiative decay
$\phi\rightarrow\pi^{0}\eta\gamma$ \cite{R9} in our calculation of
this decay rate, and this way we obtain a quadric equation for the
coupling constant $g_{\phi a_0\gamma}$ in Model I and another
quadric equation for the coupling constant $g_{a_0K^+K^-}$ in
Model II. In these quadric equations the coefficient of quadric
term results from the $a_0$ meson amplitude contribution, and the
coefficient of the linear term from the interference of the $a_0$
meson amplitude with the vector meson dominance and the loop
amplitudes. We then predict and  study the invariant mass
distribution
$dBR/dM_{\pi^0\eta}=(M_{\pi^0\eta}/M_{\phi})dBR/dE_\gamma$ for the
radiative decay $\phi\rightarrow\pi^{0}\eta\gamma$ in our
phenomenological approach using the values of coupling constants
$g_{\phi a_0\gamma}$ in Model I and $g_{a_0K^+K^-}$ in Model II
that we obtain and compare our results with the experimental
invariant $\pi^0\eta$ mass spectrum for the decay
$\phi\rightarrow\pi^{0}\eta\gamma$ \cite{R7}.

In Model I, we obtain for the coupling constant $g_{\phi
a_0\gamma}$ the values $g_{\phi a_0\gamma}=(0.24\pm 0.06)$ and
$g_{\phi a_0\gamma}=(-1.3\pm 0.3)$. In Fig. 3 we plot the
distribution $dBR/dM_{\pi^0\eta}$ choosing the coupling constant
$g_{\phi a_0\gamma}=(0.24\pm 0.06)$ in which we also indicate the
contributions coming from the different reactions shown
diagramatically in Fig. 1, as well as the contribution of the
total amplitude which includes the interference term as well. Our
Model gives a reasonable prediction for the spectrum over most of
the range of the invariant mass $M_{\pi^0\eta}$ except in its
higher part where the expected enhancement due to the contribution
of $a_0$ resonance is not produced. The distribution
$dBR/dM_{\pi^0\eta}$ we obtain for the other root, that is for
$g_{\phi a_0\gamma}=(-1.3\pm 0.3)$, is even poorer which we do not
show. We obtain the branching ratio for the decay $\phi\rightarrow
a_0\gamma$ using for the coupling constant $g_{\phi
a_0\gamma}=(0.24\pm 0.06)$ as $BR(\phi\rightarrow
a_0\gamma)=(0.2\pm 0.1)\times 10^{-5}$. However, because of the
fact that Model I does not produce a satisfactory description of
the experimental invariant $M_{\pi^0\eta}$ mass spectrum for the
decay $\phi\rightarrow\pi^{0}\eta\gamma$, we cannot consider the
value of the coupling constant $g_{\phi a_0\gamma}=(0.24\pm 0.06)$
and the resulting branching ratio $BR(\phi\rightarrow
a_0\gamma)=(0.2\pm 0.1)\times 10^{-5}$ too seriously. In this
connection, we like to note that the coupling constant $g_{\phi
a_0\gamma}$ has been calculated \cite{R21} employing QCD sum rules
and utilizing $\omega\phi$-mixing by studying the three point
$\phi a_0\gamma$-correlation function as $g_{\phi
a_0\gamma}=(0.11\pm 0.03)$. Furthermore, the photoproduction cross
section of $\rho^0$ mesons on photon targets near threshold has
been shown \cite{R22} to be mainly given by $\sigma$-exchange, and
assuming vector meson dominance of the electromagnetic current the
value of the coupling constant $g_{\rho\sigma\gamma}=2.71$ was
deduced. In the study of the structure of the $\phi$ meson
photoproduction amplitude on nucleons near threshold based on the
one-meson exchange and Pomeron-exchange mechanism \cite{R23}, this
value of the coupling constant $g_{\rho\sigma\gamma}$ was used to
calculate the coupling constant $g_{\phi a_0\gamma}$ by invoking
unitary symmetry arguments as $\mid g_{\phi a_0\gamma}\mid=0.16$
by assuming that $\sigma$, $f_0$, and $a_0$ are numbers of a
unitary nonet, which is not without problems. In the light of our
discussion of Model I and the resulting poor invariant mass
spectrum of the decay $\phi\rightarrow\pi^{0}\eta\gamma$, it may
be argued that all these values for the coupling constant $g_{\phi
a_0\gamma}$ should be considered with some caution.

We follow the same procedure in Model II, and obtain the coupling
constant $g_{a_0 K^+K^-}$ as  $g_{a_0 K^+K^-}=(-1.5\pm 0.3)$ GeV
and $g_{a_0 K^+K^-}=(3.0\pm 0.4)$ GeV utilizing the experimental
value of the $\phi\rightarrow\pi^{0}\eta\gamma$ decay rate . We
then plot the resulting invariant mass distribution for the decay
$\phi\rightarrow\pi^{0}\eta\gamma$ and compare it with the
experimental result \cite{R7}. In Fig. 4 and in Fig. 5 we plot the
distribution $dBR/dM_{\pi^0\eta}$  for the radiative decay
$\phi\rightarrow\pi^{0}\eta\gamma$ in our phenomenological
approach choosing coupling constant $g_{a_0 K^+K^-}=-1.5$ GeV and
$g_{a_0 K^+K^-}=3.0$ GeV, respectively, as a function of the
invariant mass $M_{\pi^0\eta}$ of the $\pi^0\eta$ system. In these
figures, as before, we also indicate the contributions coming from
the different reactions
$\phi\rightarrow\rho^0\pi^{0}\rightarrow\pi^0\eta\gamma$,
$\phi\rightarrow K^+K^-\gamma\rightarrow\pi^0\eta\gamma$, and
$\phi\rightarrow a_0\gamma\rightarrow\pi^0\eta\gamma$ in Model II
shown diagramatically in Fig. 2, as well as the contribution of
the total amplitude which includes the interference terms as well.
We note that for $g_{a_0 K^+K^-}=-1.5$ GeV our prediction for the
invariant mass spectrum is in good agreement with the experimental
result and not only the overall shape and feature but also the
enhancement due to the contribution of the $a_0$ resonance is well
produced. Therefore, from the analysis of the spectrum obtained
with the coupling constants $g_{a_0 K^+K^-}=-1.5$ GeV and $g_{a_0
K^+K^-}=3.0$ GeV in Fig. 4 and 5, respectively, we may decide in
favour of the value $g_{a_0K^+K^-}=-1.5$ GeV, and we may state
that the experimental data within the framework of our
phenomenological approach in Model II, suggest for the coupling
constant $g_{a_0K^+K^-}$ the value $g_{a_0K^+K^-}=(-1.5\pm 0.3)$
GeV. Furthermore, we note that Model II provides a better way as
compared to Model I in order to incorporate the $a_0$ meson into
the mechanism of the $\phi\rightarrow\pi^{0}\eta\gamma$ decay, and
thus our result gives further support for the approach in which
$a_0$ meson state arises as a dynamical state and for the two-step
mechanism for its coupling to $\phi$ meson with intermediate
$K\bar{K}$ states. In a previous work \cite{R24}, we studied the
radiative $\phi\rightarrow\pi^{0}\pi^0\gamma$ decay also within
the framework of a phenomenological approach in which we
considered the contributions of $\sigma$ meson, $\rho$ meson and
$f_0(980)$ meson. We analyzed the interference effects between
different contributions using the experimental results of SND
Collaboration. In that analysis, the coupling constant
$g_{f_0K^+K^-}$ appeared with positive sign, that is
$g_{f_0K^+K^-}>0$. In the present analysis, we obtained the
coupling constant $g_{a_0K^+K^-}$ as $g_{a_0K^+K^-}<0$. We like to
note that this result about the relative phase between
$g_{f_0K^+K^-}$ and $g_{a_0K^+K^-}$ is consistent with the result
obtained in the $q^2\bar{q}^2$ model where
$g_{f_0K^+K^-}=-g_{a_0K^+K^-}$ \cite{R3,R4,R25}.

We then calculate the decay rate $\Gamma(\phi\rightarrow
a_0\gamma)$ by assuming the two-step one-loop mechanism
$\phi\rightarrow K^+K^-\rightarrow a_0\gamma$ for the decay
$\phi\rightarrow a_0\gamma$. We show this mechanism
diagramatically in Fig. 6. The decay rate that follows from these
diagrams is given by
\begin{eqnarray}\label{e20}
\Gamma (\phi\rightarrow a_0\gamma)= \frac{\alpha g_{\phi K^+K^-}^2
g_{a_0 KK }^2}{3(2\pi)^4}\frac{E_\gamma}{M_\phi^2}\mid
(a-b)I(a,b)\mid^2 ~~.
\end{eqnarray}
where $a=M_{\phi}^{2}/M_{K}^{2}$, $b=M_{a_0}^2/M_{K}^{2}$, and the
loop function I(a, b) is defined in Eq. 11. We use the coupling
constant $g_{a_0 KK}=(-1.5\pm 0.3)$ GeV and then obtain the decay
rate for $\Gamma(\phi\rightarrow a_0\gamma)$ decay as
$\Gamma(\phi\rightarrow a_0\gamma)=(0.51\pm 0.09)$ KeV and the
branching ratio as $BR(\phi\rightarrow a_0\gamma)=(1.1\pm
0.2)\times 10^{-4}$. This result when combined with the
experimental value $BR(\phi\rightarrow f_0\gamma)=(3.4\pm
0.4)\times 10^{-4}$ gives the ratio $\Gamma(\phi\rightarrow
f_0\gamma)/\Gamma(\phi\rightarrow a_0\gamma)=(3.0\pm 0.9)$, which
is still larger than the theoretical value of 1.The value
$BR(\phi\rightarrow a_0\gamma)=(1.1\pm 0.2)\times 10^{-4}$ we
obtain for the branching ratio of the decay $\phi\rightarrow
a_0\gamma$, although not in disagreement, is somewhat larger than
the result of Achasov et al. which is $BR(\phi\rightarrow
a_0\gamma)=(0.88\pm 0.17)\times 10^{-4}$ \cite{R7}. Consequently,
we obtain a smaller value for the experimental ratio
$\Gamma(\phi\rightarrow f_0\gamma)/\Gamma(\phi\rightarrow
a_0\gamma)$ than the previous result which is
$\Gamma(\phi\rightarrow f_0\gamma)/\Gamma(\phi\rightarrow
a_0\gamma)=(3.8\pm 1.2)$. We like to note, however, that Achasov
et al \cite{R7} assumed that $a_0\gamma$ intermediate state
dominates the $\phi\rightarrow \pi^0\eta\gamma$ decay and they
neglected the contributions coming from other intermediate states
in their analysis. On the other hand, in our analysis we include
the contributions coming not only from $a_0\gamma$ intermediate
state but also from $\rho^0\pi^0$ and $K^+K^-$ intermediate states
as well as from their interference.

In order to obtain the branching ratio for $\phi\rightarrow
a_0\gamma$ decay utilizing the experimental value of the decay
rate and the experimental invariant $\pi^0\eta$ mass spectrum for
the $\phi\rightarrow\pi^{0}\eta\gamma$ decay, we use a
phenomenological approach. In our analysis, we employ point-like
effective field theory and we deduce from the experimental data
the coupling  constant of point-like $a_0$ meson coupled to
point-like K meson. The coupling constant that we obtain $g_{a_0
K^+K^-}=(-1.5\pm 0.3)$ GeV results in the branching ratio
$BR(\phi\rightarrow a_0\gamma)=(1.1\pm 0.2)\times 10^{-4}$ for
$\phi\rightarrow a_0\gamma$ decay using one loop $\phi\rightarrow
K^+ K^-\rightarrow a_0\gamma$ mechanism. We can, therefore,
suggest that the branching ratio for the decay $\phi\rightarrow
a_0\gamma$ that is used in the literature should be revised as
$BR(\phi\rightarrow a_0\gamma)=(1.1\pm 0.2)\times 10^{-4}\times
F^2(R)$. We can also assert that our analysis suggests a lower
value $(3.0\pm 0.9)$ for the ratio $\Gamma(\phi\rightarrow
f_0\gamma)/\Gamma(\phi\rightarrow a_0\gamma)$ than used
previously.

In a very recent paper \cite{R26} the KLOE Collaboration presented
the result of their experimental study of
$\phi\rightarrow\pi^{0}\eta\gamma$ decay. In their analysis, they
conclude that $\phi\rightarrow\pi^{0}\eta\gamma$ decay is
dominated by the process $\phi\rightarrow a_0\gamma$ and from a
fit to $\pi^0\eta$ invariant mass spectrum they find the branching
ratio $BR(\phi\rightarrow a_0\gamma)=(7.4\pm 0.7)\times 10^{-5}$
for the $\phi\rightarrow a_0\gamma$ decay. In view of this
finding, our result $BR(\phi\rightarrow a_0\gamma)=(1.1\pm
0.2)\times 10^{-4}\times F^2(R)$ implies that $F^2(R)=(0.7\pm
0.2)$. This value supports the view that the structure of $a_0$
meson is a combination of a $K\bar{K}$ molecule with a compact
$q^2\bar{q}^2$ core \cite{R27}.

\newpage
\begin{figure}
\epsfig{figure=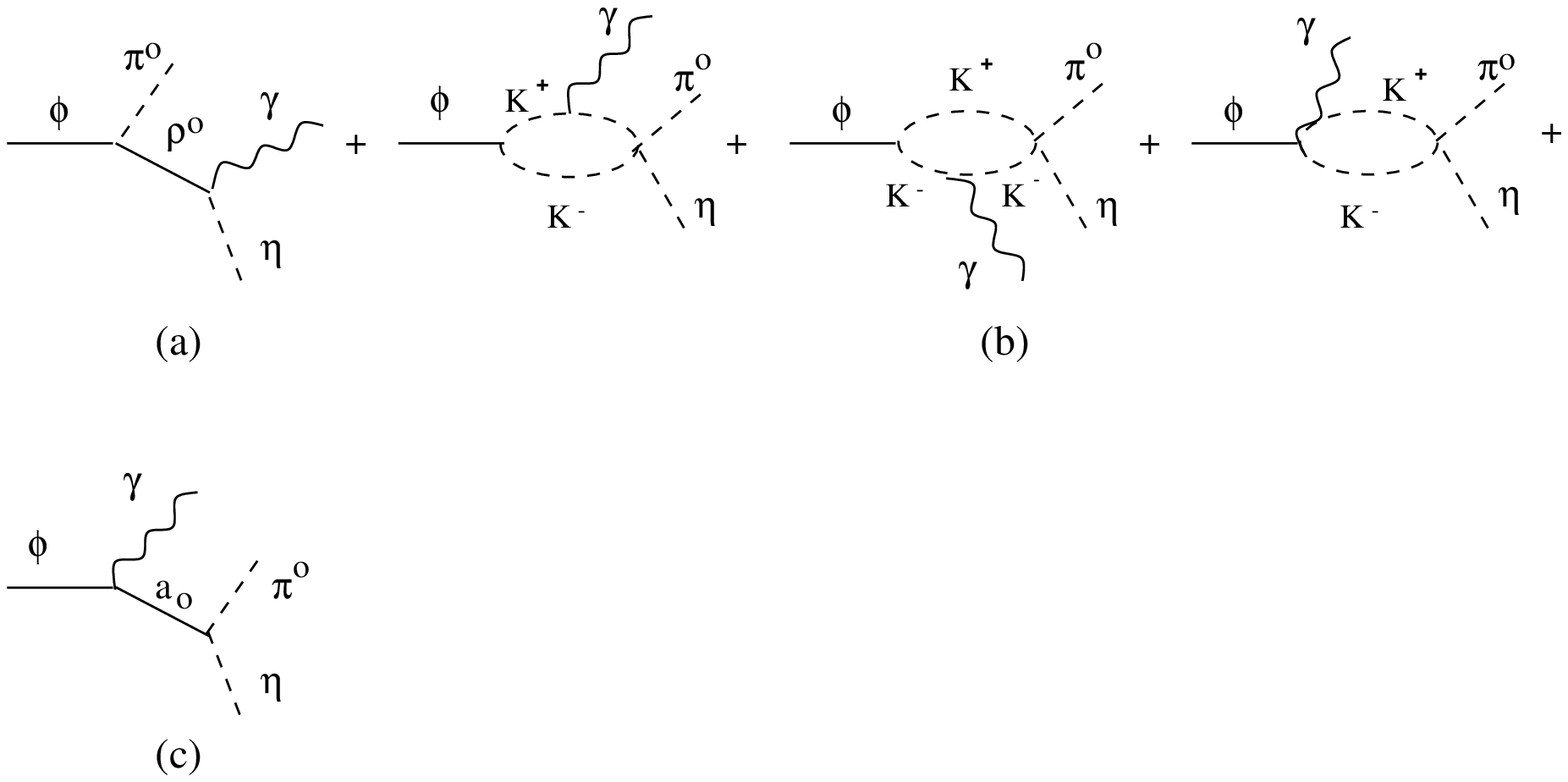,height=15cm, angle=270}\vspace*{1.0cm}
\caption{Diagrams for the decay $\phi\rightarrow \pi^0\eta\gamma$
in Model I.}
\end{figure}

\begin{figure}
\vspace*{1.0cm} \epsfig{figure=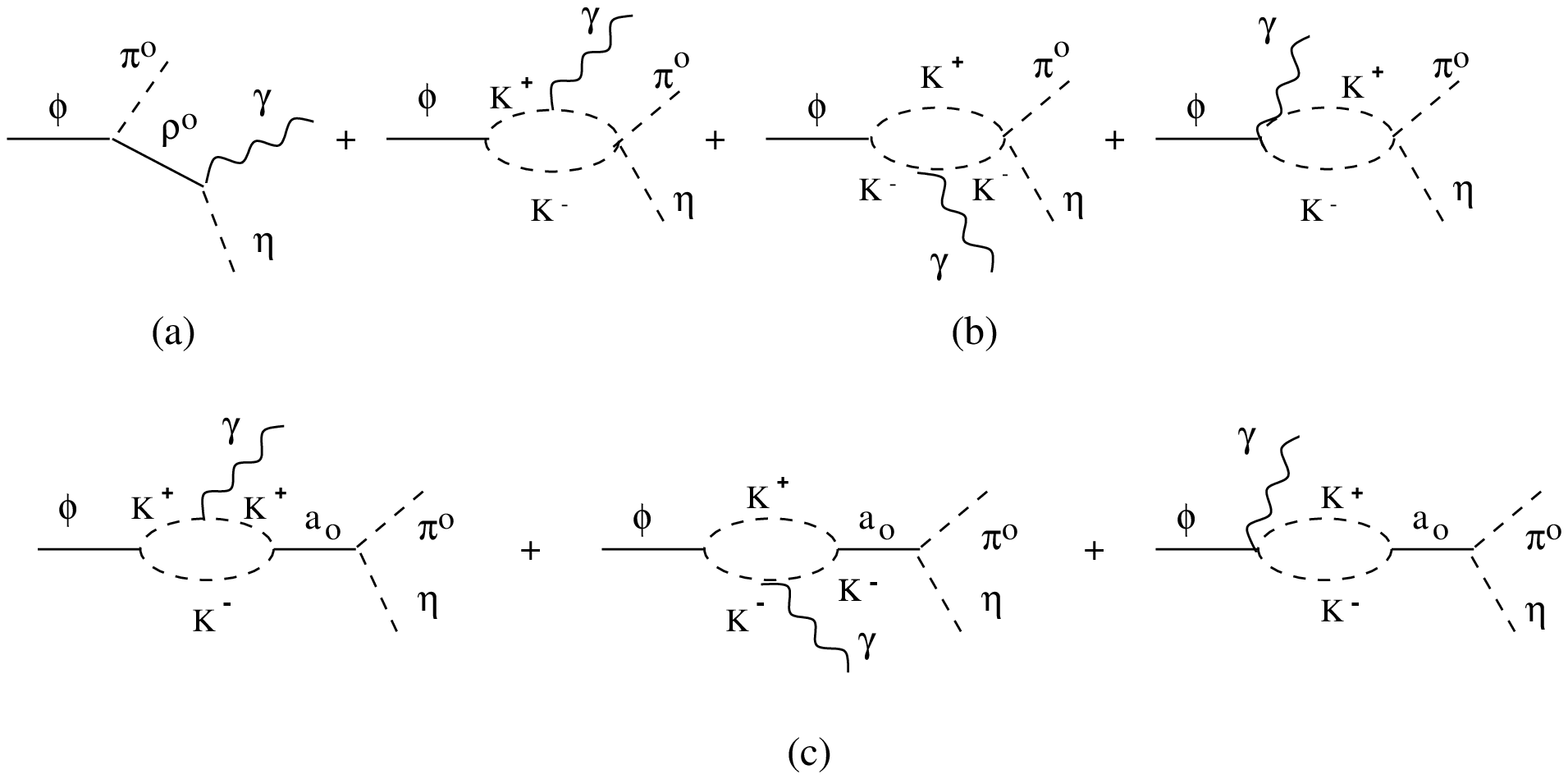,height=15cm,
angle=270}\vspace*{1.0cm} \caption{Diagrams for the decay
$\phi\rightarrow \pi^0\eta\gamma$ in Model II.}
\end{figure}

\newpage

\begin{figure}
\epsfig{figure=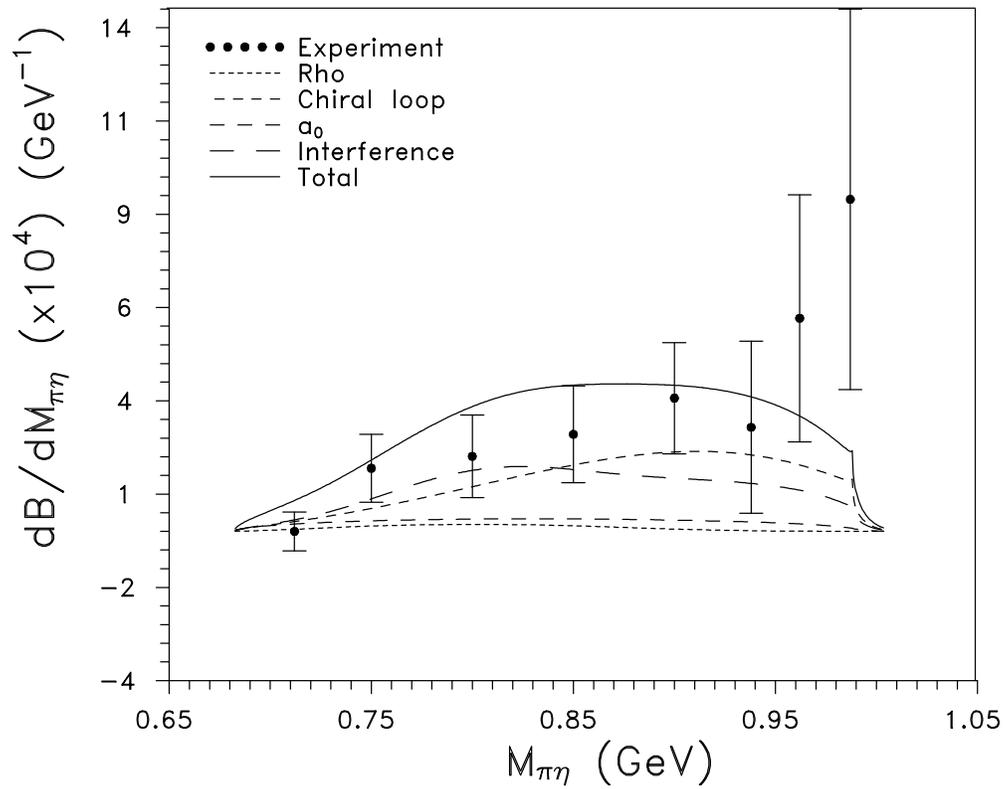,height=20cm}\vspace*{-3.0cm} \caption{The
$\pi^0\eta$ invariant mass spectrum for the decay
$\phi\rightarrow\pi^{0}\eta\gamma$ in Model I. The contributions
of different terms are indicated.}
\end{figure}

\newpage

\begin{figure}
\epsfig{figure=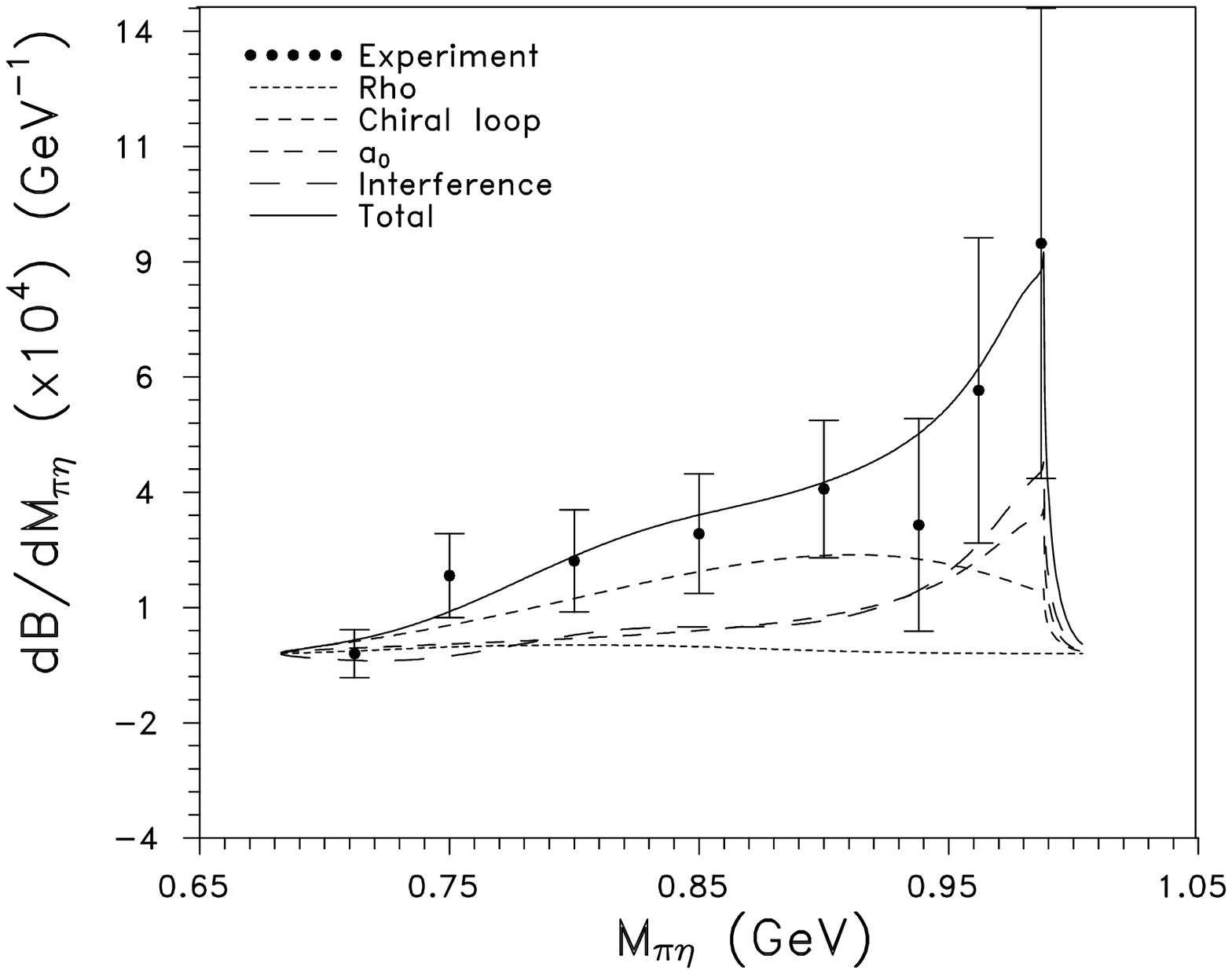,height=20cm} \vspace*{-3.0cm} \caption{The
$\pi^0\eta$ invariant mass spectrum for the decay
$\phi\rightarrow\pi^{0}\eta\gamma$ for g$_{a_0K^+K^-}=-1.5$ GeV in
Model II. The contributions of different terms are indicated.}
\end{figure}

\newpage

\begin{figure}
\epsfig{figure=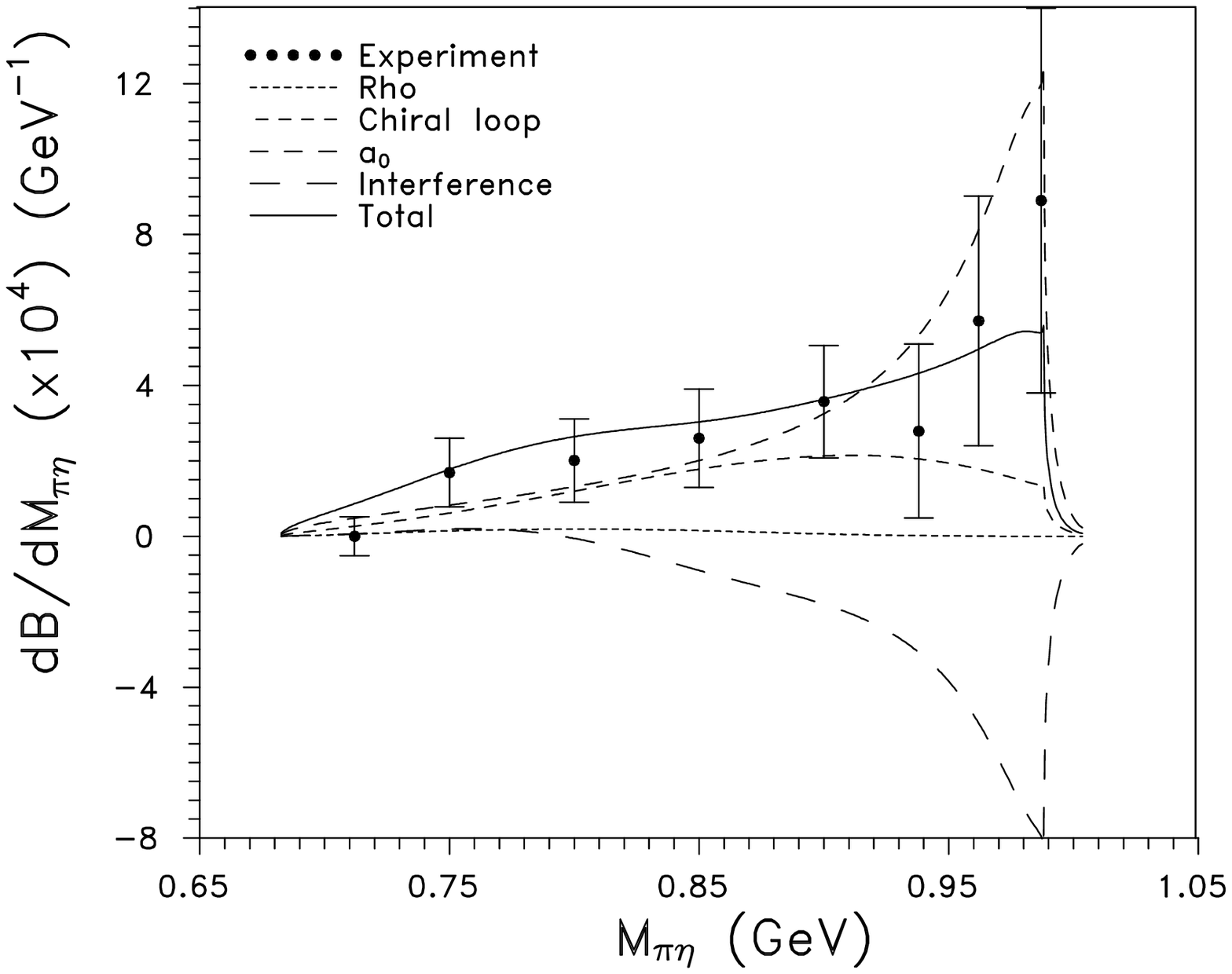,height=20cm}\vspace*{-3.0cm} \caption{The
$\pi^0\eta$ invariant mass spectrum for the decay
$\phi\rightarrow\pi^{0}\eta\gamma$ for g$_{a_0K^+K^-}=3.0$ GeV in
Model II. The contributions of different terms are indicated.}
\end{figure}

\newpage

\begin{figure}
\vspace*{2.0cm} \epsfig{figure=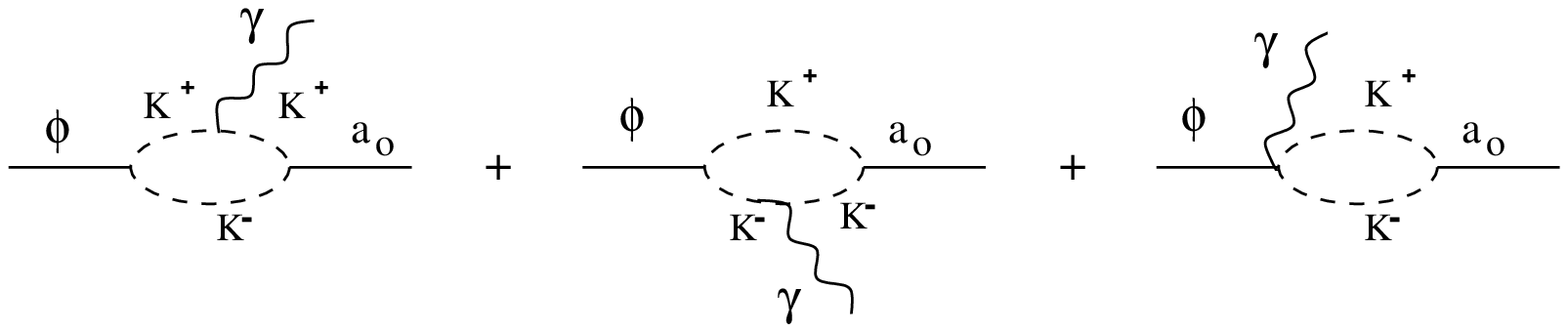,height=15cm,
angle=270}\vspace*{2.0cm} \caption{ Diagrams for the decay
$\phi\rightarrow a_0\gamma$.}
\end{figure}

\end{document}